\pdfoutput=1

\documentclass[11pt]{article}

\usepackage[preprint]{acl}

\usepackage{times}
\usepackage{latexsym}

\usepackage[T1]{fontenc}

\usepackage[utf8]{inputenc}

\usepackage{microtype}

\usepackage{inconsolata}

\usepackage{graphicx}
\usepackage{tabularx}
\usepackage{booktabs}
\usepackage{multirow}
\usepackage{amsmath}
\usepackage{cleveref}
\usepackage[most]{tcolorbox}
\usepackage{lipsum} 

\tcbset{
    userstyle/.style={
        enhanced,
        colback=white,
        colframe=black,
        colbacktitle=gray!20,
        coltitle=black,
        rounded corners,
        sharp corners=north,
        boxrule=0.5pt,
        drop shadow=black!50!white,
        attach boxed title to top left={
            xshift=-2mm,
            yshift=-2mm
        },
        boxed title style={
            rounded corners,
            size=small,
            colback=gray!20
        }
    },
    replystyleg/.style={
        enhanced,
        colback=green!15,
        colframe=black,
        colbacktitle=green!30,
        coltitle=black,
        boxrule=0.5pt,
        drop shadow=black!50!white,
        rounded corners,
        sharp corners=north,
        attach boxed title to top right={
            xshift=-2mm,
            yshift=-2mm
        },
        boxed title style={
            rounded corners,
            size=small,
            colback=green!40
        }
    },
}

\newtcolorbox{Swahiliquery}[1][]{
    userstyle,
    title=Prompt (Swahili),
    #1
}

\newtcolorbox{Indonesianquery}[1][]{
    userstyle,
    title=Prompt (Indonesian),
    #1
}

\newtcolorbox{Swahilireply-g}[1][]{
    replystyleg,
    title=Response (meaningless repetition),
    #1
}

\newtcolorbox{Indonesianreply-g}[1][]{
    replystyleg,
    title=Response (irrelevant answer),
    #1
}

\title{Beyond the Tip of Efficiency: Uncovering the Submerged Threats of \\ Jailbreak Attacks in Small Language Models}

\author{
    \begin{tabular}{c}
    Sibo Yi$^{1}$ \quad Tianshuo Cong$^{1,4}$\quad Xinlei He$^3$\quad
    Qi Li$^{1,2}$\quad Jiaxing Song$^{1,2}$\footnotemark[1] 
    \end{tabular}
    \\
    $^1$ Tsinghua University \ 
    $^2$ Zhongguancun Laboratory \\
    $^3$ The Hong Kong University of Science and Technology (Guangzhou) \\
    $^4$ Shandong Key Laboratory of  Artificial Intelligence Security
    \\
    \texttt{ysb23@mails.tsinghua.edu.cn},
    \texttt{xinleihe@hkust-gz.edu.cn},
    \\
    \texttt{\{congtianshuo,qli01,jxsong\}@tsinghua.edu.cn}
}

\begin{document}
\maketitle
\begin{abstract}
Small language models (SLMs) have become increasingly prominent in the deployment on edge devices due to their high efficiency and low computational cost. 
While researchers continue to advance the capabilities of SLMs through innovative training strategies and model compression techniques, the security risks of SLMs have received considerably less attention compared to large language models (LLMs).
To fill this gap, we provide a comprehensive empirical study to evaluate the security performance of 13 state-of-the-art SLMs under various jailbreak attacks. 
Our experiments demonstrate that most SLMs are quite susceptible to existing jailbreak attacks, while some of them are even vulnerable to direct harmful prompts.
To address the safety concerns, we evaluate several representative defense methods and demonstrate their effectiveness in enhancing the security of SLMs. 
We further analyze the potential security degradation caused by different SLM techniques including architecture compression, quantization, knowledge distillation, and so on. 
We expect that our research can highlight the security challenges of SLMs and provide valuable insights to future work in developing more robust and secure SLMs.   
\end{abstract}

\begingroup
\renewcommand{\thefootnote}{\fnsymbol{footnote}}

\footnotetext[1]{Corresponding author.}
\endgroup

\section{Introduction}

Large language models (LLMs), such as ChatGPT~\cite{GPT20, ouyang2022training,achiam2023gpt} and Llama series~\cite{LLAMA23,dubey2024llama}, have demonstrated revolutionary performance in a spectrum of text generation tasks. 
As a fundamental principle for guiding the development of LLMs, the scaling law~\cite{kaplan2020scaling} highlights the strong correlation between the performance and scale of LLMs. 
However, as LLMs evolve to encompass hundreds and even thousands of billions of parameters, their development imposes expensive demands on computational resources and high-quality data for pre-training. 
Consequently, LLMs are typically confined to deployment on GPU clusters and cloud environments, posing significant challenges for wide adoption on edge devices such as smartphones, laptops, autonomous vehicles, and wearables.

Recently, small language models (SLMs) have attracted significant attention from the academic community for their efficiency and remarkable performance in various tasks~\cite{lu2024small, vannguyen2024surveys}.
On platforms like Hugging Face, SLM collections such as Llama-3.2~\cite{dubey2024llama}, MiniCPM ~\cite{hu2024minicpm} and Phi~\cite{abdin2024phi} have gained considerable popularity among researchers and achieved top-tier download rates. 
Different from LLMs, SLMs typically consist of only a few billion parameters, requiring significantly less training data and computational cost for deployment. 

However, unlike LLMs that benefit from extensive datasets and robust alignment strategies, it is challenging for SLMs to balance between generation capabilities and security, which makes them more vulnerable to jailbreak attacks. 
Among these, one of the most serious threats is referred to as jailbreak. 
By creating malicious prompts to induce target LLMs to generate harmful responses, jailbreak has emerged as a critical security concern in the development of LLMs~\cite{yi2024jailbreak, Yao2024, GAAPP23, He2025AISecuritySurvey}.
Moreover, certain jailbreak techniques that can bypass the security boundary of LLMs or VLMs have demonstrated strong transferability to other models~\cite{ZWKF23, zhang2025fc, gong2025figstep}, which presents potential threats to all generative models including SLMs. 

Although security concerns regarding SLMs have become an increasingly important issue, there still remains a substantial gap in exploring and understanding the security boundary of SLMs. 
In this paper, we collect representative malicious datasets, jailbreak attack methods, and defense methods to conduct adversarial experiments on numerous state-of-the-art SLMs, thereby revealing existing security vulnerabilities of SLMs and exploring corresponding mitigation strategies. 
Furthermore, we take insight into the security degradation of SLMs and discuss some potential factors.
In summary, we make the following contributions:

\begin{itemize}
    \item {We conduct extensive experiments to reveal the security vulnerabilities of SLMs under different jailbreak attacks. Especially, The results demonstrate that most SLMs are more susceptible to jailbreak attacks compared to LLMs.}
    \item {We evaluate the effectiveness of existing defense methods on SLMs. The results show that these methods are significantly adapted to SLMs to enhance their resilience against jailbreak attacks.}
    \item {We discuss and analyze various underlying factors that may lead to the security degradation of SLMs, including inadequate safety alignment, biased knowledge distillation, parameter sharing, and quantization techniques.}
\end{itemize}

\section{Related Work}
\subsection{Jailbreak Attacks}
Jailbreak attacks, which transform harmful queries like ``How to make a bomb'' into more sophisticated prompts to deceive target models to generate toxic output, can be mainly classified into two categories: white-box methods and black-box methods. 

White-box methods generally rely on access to the internal states of LLMs to design attack strategies. 
These methods generally use gradients and logits of target LLMs as loss functions to optimize adversarial suffixes appended to malicious questions~\cite{ZWKF23, JDRS23, ZZAWBWHNS23, andriushchenko2024jailbreaking, geisler2024attacking, mangaokar2024prp}, or manipulate the output logits to enforce target LLMs to generate affirmative responses~\cite{HGXLC24, zhang2024jailbreak}.
However, white-box methods tend to generate irregular prompts that are easily detectable and cannot be optimized directly on black-box models like ChatGPT.

In contrast, black-box methods construct readable prompts in different ways and validate their effectiveness based on the responses of the target LLMs. 
Some studies employ heuristic strategies to rewrite malicious questions in other formats such as ASCII format~\cite{jiang2024artprompt}, code format~\cite{KLSGZH23, lv2024codechameleon}, encrypted format~\cite{YJWHHST24, liu2024making} and low-resource languages~\cite{DZPB24}, exploiting the insufficient safety alignment of target LLMs in these formats to bypass the defense mechanism.
Another line of research is to instruct an advanced LLM like GPT-4 to optimize jailbreak prompts by incorporating iterative refinement~\cite{jin2024guard}, genetic algorithms~\cite{LXCX23} and psychological expertise~\cite{ZLZYJS24}, or fine-tune another LLM with successful jailbreak templates to serve as an attacker to generate jailbreak prompts automatically~\cite{deng2024masterkey, ge2023mart}.
Compared with white-box methods, black-box methods can be applied to most models. For this reason, black-box methods are widely used in empirical experiments to evaluate the safety of LLMs.

To mitigate threats caused by jailbreak attacks, different defense techniques are proposed to ensure the security of LLMs~\cite{wang2024defending, xiong2024defensive, ouyang2025layer, xu2024safedecoding, gao2024shaping}. One line of work addresses the issue by detecting~\cite{inan2023llama} or perturbing~\cite{RWHP23} the jailbreak prompts to reduce the toxicity of the input, while another line of work directly enhances the robustness of LLMs by supervised fine-tuning~\cite{BSARJHZ24} or reinforcement learning from human feedback~\cite{Ouayng22}. 

\subsection{Small Language Models}
Similar to LLMs, SLMs are typically built upon decoder-only architectures, while they show diversity in implementation details such as the type of attention heads, layer numbers, dimension sizes, activation functions, and so on. 

\begin{figure*}[t]
    \centering
    \includegraphics[width=0.9\linewidth]{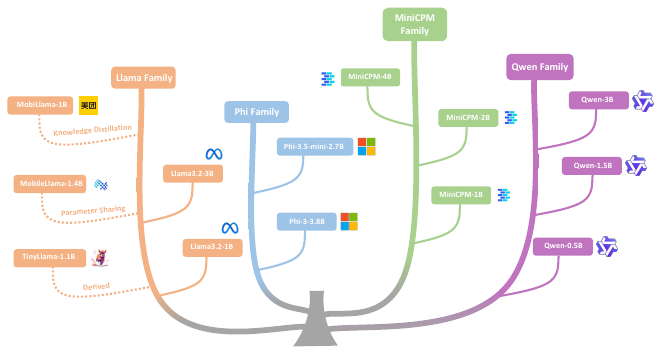}
    \caption{The family tree of the target SLMs we evaluate in our paper. The solid line represents the model is belonging to a certain family, while the dashed line indicates that the model is derived from that family with certain SLM technology.}
    \label{fig:overall}
\end{figure*}

To achieve competitive performance within the limited scale of SLMs, different model compression techniques are adopted to construct lightweight architectures efficiently. 
For instance, MobilLLaMA~\cite{thawakar2024mobillama} and MobileLLM~\cite{chu2023mobilevlm} introduce a parameter-sharing scheme in embedding blocks and attention head blocks to reduce the cost of GPU memory. 
TinyLLaMA~\cite{zhang2024tinyllama} optimizes memory load with the FlashAttention technique~\cite{dao2022flashattention}, which introduces an IO-aware attention algorithm to reduce the budget of high bandwidth memory. 
Quantization techniques, such as GPTQ~\cite{frantar2022gptq} and AWQ~\cite{lin2024awq}, can also effectively reduce memory loads by compressing the size parameters from 16 bits to 8 bits or even 4 bits.  
In model collections such as Llama 3~\cite{LLAMA23}, Qwen~\cite{bai2023qwen}, and MiniCPM~\cite{hu2024minicpm}, SLMs are generally designed and pre-trained following LLMs in the same family. 
Additionally, during the training phase, knowledge distillation techniques are widely used to derive performance from teacher LLMs to student SLMs, as models in the same family generally share similar tokenizers and architecture.

Recent research has demonstrated that SLMs can achieve comparable performance in some reasoning tasks, and can even outperform LLMs in specific scenarios~\cite{lu2024small}. 
Our study fills a gap in evaluating the security of SLMs from another perspective. In the following sections, we will demonstrate the security differences between various SLMs and delve into their underlying causes.

\section{Experiment Setups}

\subsection{Target Models}
We collect 16 state-of-the-art models to provide a comprehensive view of their security differences, including 13 SLMs below 4B size and 3 LLMs above 7B size.

For LLMs, we include Llama2-7B~\cite{Llama2}, Llama3-8B~\cite{LLAMA23}, and DeepSeek-R1-Distill-Llama-8B~\cite{guo2025deepseek} for evaluation. 
Notably, DeepSeek-R1-Distill-Llama-8B is constructed by distilling the reasoning patterns from DeepSeek-R1 to Llama3-8B.
The controlled comparison enables us gain some valuable insights of the influences of distillation techniques to SLM security.
For SLMs, we include 13 models from advanced research organizations and individual developers. Specifically, they are as follows:

\begin{itemize}
    \item{\textbf{Llama Family}. Llama family is developed by Meta AI as one of the most popular model series. For our study, we select two models from the Llama 3.2 collection with parameter sizes of 1B and 3B. Furthermore, we include three additional SLMs that are initialized from Llama and processed with certain model compression techniques. These models are MobilLLaMA~\cite{thawakar2024mobillama}, MobileLLM~\cite{chu2023mobilevlm}, and TinyLLaMA~\cite{zhang2024tinyllama}.}
    
    \item{\textbf{Phi Family}. Phi family~\cite{gunasekar2023textbooks} is developed by Microsoft focusing on designing lightweight SLMs with exceptional performance. We select Phi-3-mini-4k-instruct in 3.8B size and Phi-3.5-mini-instruct size in 2.7B size for evaluation.}
    
    \item{\textbf{MiniCPM Family}. MiniCPM family~\cite{hu2024minicpm} is developed by OpenBMB and mainly consists of SLMs in different versions. We select MiniCPM-1B-sft-bf16, MiniCPM-2B-sft-bf16 and MiniCPM-4B for evaluation.}
    
    \item{\textbf{Qwen Family}. Qwen family~\cite{bai2023qwen} is built by Alibaba Cloud which has released a spectrum of LLMs ranging from 0.5B to 72B sizes. We select Qwen2.5-0.5B-Instruct, Qwen2.5-1.5B-Instruct, and Qwen2.5-3B-Instruct for evaluation.}
\end{itemize}

\subsection{Attack Methods}
Our research firstly examines the influences of direct attacks against SLMs, which leverage straightforward harmful queries such as ``How to make a bomb'' to probe the target models directly. 
We collect 5 datasets that contain such harmful questions across various illegal and unethical dimensions.

\begin{itemize}
    \item{\textbf{Advbench}. Advbench~\cite{ZWKF23} is a harmful dataset consisting of 500 harmful strings and 500 harmful behaviors. The former focuses on eliciting specific harmful responses from target LLMs, while the latter aims at provoking the models into exhibiting harmful behavior as much as possible.}
    
    \item{\textbf{DAN}. DAN~\cite{SCBSZ23} provides a forbidden question set spanning 13 restricted scenarios. The dataset is primarily sourced from online platforms and publicly available datasets.}
    
    \item{\textbf{maliciousInstruct}. MaliciousInstruct~\cite{HGXLC24} consists of 100 harmful questions with 10 malicious intentions, which are mostly generated by  ChatGPT and then revised manually.}
    
    \item{\textbf{StrongREJECT}. StrongREJECT~\cite{souly2024strongreject} offers 313 harmful questions that cover forbidden scenarios from different AI usage policies. The majority of the dataset is written manually, while the remaining portion is sourced from LLMs and other open-source datasets.}
    
    \item{\textbf{XSTEST}. XSTEST~\cite{RKVABH23} contains both safe and unsafe questions to assess the exaggerated safety behaviors of models. We extract the 200 harmful questions from the dataset for our experiments.}
\end{itemize}

Furthermore, to explore and understand the safety boundary of SLMs more clearly, we conduct a thorough investigation into existing jailbreak attacks and select 5 representative methods for evaluation. 
These methods span across different categories of jailbreak attacks and have demonstrated excellent effectiveness against LLMs in previous studies, thereby enabling a comprehensive and reliable evaluation of the robustness of SLMs.

\begin{itemize}
    \item{\textbf{GCG}. Greedy Coordinate Gradient (GCG)~\cite{ZWKF23} is a gradient-based attack that initializes an adversarial suffix appended to the malicious question and optimizes it by gradient-based search to maximize the probability of affirmative responses. Although the optimization of jailbreak prompts is constrained to white-box models, they demonstrate strong transferability to other black-box models.}
    
    \item{\textbf{ArtPrompt}. ArtPrompt~\cite{jiang2024artprompt} is an ASCII-based attack that leverages the poor performance of LLMs in recognizing ASCII art to bypass defense mechanisms. Specifically, ArtPrompt utilizes LLMs like GPT-4 to recognize the malicious word in the prompt and visually encodes it with ASCII characters, combining the text prompt and the word in ASCII art to jailbreak.}
    
    \item{\textbf{DeepInception}. DeepInception~\cite{LZZYLH23} is a template-based attack that embeds malicious questions into virtual scenarios. Given a harmful question, DeepIncetion constructs a multi-layer scene with different characters and induces the target LLMs to complement the story step by step, thus generating harmful content in responses.}
    
    \item{\textbf{AutoDAN}. AutoDAN is a genetic algorithm-based attack that refines jailbreak prompts iteratively to identify the optimal solution. Specifically, AutoDAN randomly initializes the original jailbreak population and performs word-level or sentence-level modifications to produce offspring. The new generation is subsequently evaluated by LLMs to gain fitness and repeat the generation process until the jailbreak succeeds. }
    
    \item{\textbf{Multilingual Attack}. Multilingual attack~\cite{DZPB24}} exploits the weakness of the safety alignment in low-resource languages to conduct jailbreak attacks. The method translates harmful questions into multiple languages and shows that questions in low-resource languages demonstrate a high attack success rate.
\end{itemize}

Notably, the datasets are all sourced from official resources to guarantee the reliability and robustness of the experimental results. For jailbreak methods, we follow the official implementation and use the best parameter settings as reported in the original papers. 

\subsection{Defense Methods}
In addition to examining the effectiveness of different jailbreak attacks against SLMs, we have also explored potential defense strategies to mitigate these threats and enhance the robustness of SLMs. 

Given the extensive set of SLMs and attack methods involved in our experimental design, we deliberately prioritized two key criteria when selecting defense methods for evaluation.
On the one hand, the defense methods should be applicable to diverse SLMs under different jailbreak attacks without requiring attack-specific or model-specific modification. This ensures the evaluation of defense methods can be conducted under unified experimental settings, thereby guaranteeing the fairness and reliability of the results.
On the other hand, the defense method should not add a substantial burden to the efficient execution of SLMs. Otherwise, the defense would be difficult to deploy in real-world applications.

Finally, we collect 3 defense methods with the required high transferability and efficiency.

\begin{itemize}
    \item{\textbf{Llama-Guard-3}. Llama-Guard-3~\cite{inan2023llama} is fine-tuned from Llama-3 to detect unsafe content within user prompts and LLM responses. In our study, we instruct Llama-Guard-3-8B to detect jailbreak prompts and filter user inputs that are judged to be unsafe.}
    
    \item{\textbf{SmoothLLM}. SmoothLLM~\cite{RWHP23} is a perturbation-based method that can mitigate malicious content in user prompts. For each prompt, SmoothLLM generates multiple copies with character-level perturbations applied to them and aggregates the responses of the target LLM to these copies to produce the final response.}

    \item{\textbf{Backtranslation}. Backtranslation~\cite{wang2024defending}} is a prompt-based defense method that can reveal the implicit malice of jailbreak attacks. Given the response of a jailbreak attack, Backtranslation instructs an LLM to recover the explicit malicious question from the response and input it again into the target model to examine whether the question is rejected.   
\end{itemize}

\subsection{Evaluation Metrics}
We use \textbf{Attack Success Rate (ASR)} as the primary metric in our experiments, which is widely used in existing evaluation toolkits~\cite{ran2024jailbreakeval, mazeika2024harmbench} to identify the effectiveness of jailbreak attacks.
Formally, ASR can be defined as 

\begin{figure*}[t]
    \centering
    \includegraphics[width=0.9\linewidth]{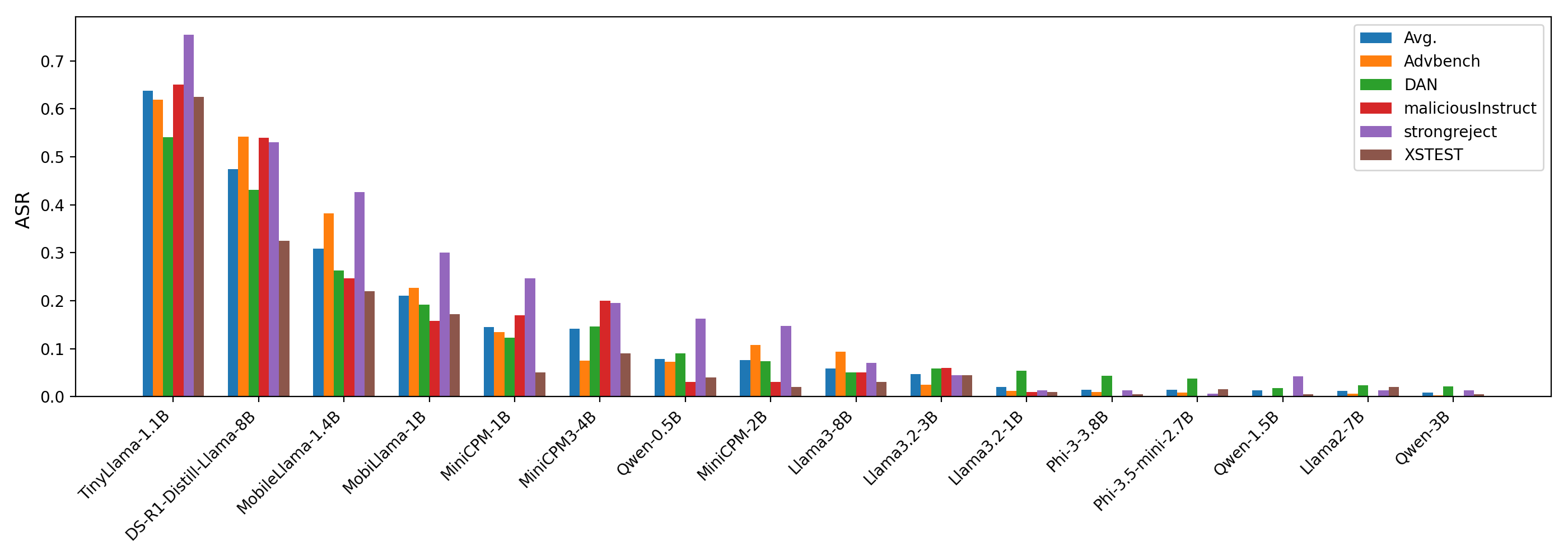}
    \caption{The security performance of the 13 SLMs and 3 LLMs under direct attacks. The security performance of target models is ranked in descending order based on the average ASR.}
    \label{fig:result1}
\end{figure*}

\begin{equation}
ASR = \frac{N_{success}}{N_{total}},
\label{ASR}
\end{equation}
where $N_{success}$ is the number of successfully attacked prompts and $N_{total}$ is the total number of jailbreak prompts. 
Rule-based matching and LLM evaluators are the most common methods to assess the success of a jailbreak attack. 
However, during the experiments, we have observed that target SLMs occasionally generated unexpected responses that are not related to the prompts, resulting in a noticeably inflated ASR when relying on rule-based matching. 
To address this issue, we ultimately employed Llama-Guard-3-8B as the evaluator to assess the responses of jailbreak attacks to calculate the ASR accurately.

\subsection{Experimental Settings}
We control the parameter settings consistently when generating responses from the target models to ensure the comparability of the results.
Specifically, we invoke the conversation template of target models to generate prompts without system prompts.
We disabled token sampling and adopt greedy decoding during output generation to ensure the reproducibility of the results.

\section{Main Results}

\subsection{Direct Attacks Against SLMs}
\label{sec:direct}
We first evaluate the fundamental defense capabilities of SLMs with direct harmful questions used as original prompts.
As illustrated in~\Cref{fig:result1}, experimental results indicate that when faced with direct attacks, most SLMs successfully identify the malicious intention and generate rejection responses, exhibiting reliable defense capabilities that are comparable to LLMs.
For SLM series including Llama, Phi, MiniCPM, and Qwen, the ASR is generally around or below 10\%.
In contrast, other models, including TinyLlama, MobileLlama, and MobiLlama, exhibit comparatively weaker performance in resisting harmful queries.
Furthermore, all target SLMs show transferable defensive capabilities across various harmful datasets. That is, if they perform well on one dataset, they tend to demonstrate similar performance on the remaining four datasets. 

As shown in~\Cref{fig:result1}, there exists a slight positive correlation between parameter size and the security performance of SLMs, which also means that parameter size is not the primary factor in determining the security of SLMs. 
For SLMs in the same series but different in parameter sizes, such as Qwen-1.5B and Qwen-3B, their security capabilities against direct attacks show minimal variation. 
Meanwhile, although TinyLlama-1.1B, Llama3.2-1B, and MiniCPM-1B have a similar parameter size, the ASR of direct attacks against TinyLlama-1.1B significantly exceeds the other two models.

\subsection{Jailbreak Attacks Against SLMs}
\label{sec:jailbreak}
\begin{figure*}[t]
    \centering
    \includegraphics[width=0.9\linewidth]{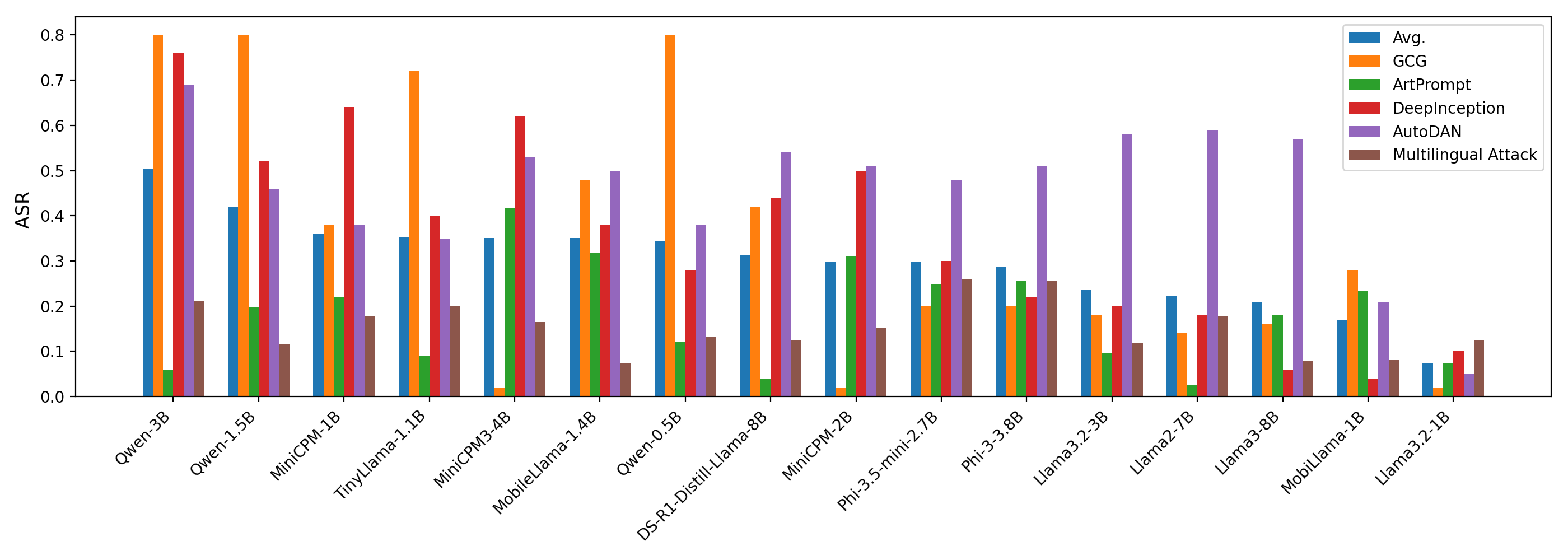}
    \caption{The security performance of the 13 SLMs and 3 LLMs under jailbreak attacks. The security performance of target models is ranked in descending order based on the average ASR.}
    \label{fig:result2}
\end{figure*}

While direct attacks can evaluate the basic robustness of SLMs, they still represent relatively simplistic attack scenarios. 
In real-world settings, adversaries tend to employ sophisticated jailbreak techniques to intentionally bypass the safety guardrails of SLMs. 
Therefore, to further examine the robustness of SLMs under more realisitic and challenging conditions, we evaluate their performance against five representative jailbreak attacks. 
The results are shown in~\Cref{fig:result2}. 
Compared with direct attacks, jailbreak attacks generally achieve higher ASR against most SLMs, with ASR on SLMs typically surpassing that on LLMs.
This suggests although most SLMs can maintain their robustness under direct attacks, they still demonstrate vulnerabilities when exposed to more sophisticated jailbreak attacks.

\begin{table*}[t]
\centering
\caption{The ASR of target models under different combinations of attacks and defenses.}
\label{tab:result3}
\resizebox{\textwidth}{!}
{
\begin{tabular}{lccccc}
\toprule
\multirow{2}{*}{Jailbreak Methods} & \multirow{2}{*}{Defenses} & \multicolumn{4}{c}{Target LLMs} \\
\cmidrule(lr){3-6}
& & TinyLlama-1.1B & Phi-3.5-mini-2.7B & MiniCPM3-1B & Qwen-3B \\
\midrule
\multirow{4}{*}{GCG} 
    & - & 0.72 & 0.20 & 0.38 & 0.80 \\
    & Llama-Guard-3 & 0 (-0.72) & 0 (-0.20) & 0 (-0.38) & 0 (-0.80) \\
    & SmoothLLM & 0.20 (-0.52) & 0.02 (-0.18) & 0.12 (-0.26) & 0.02 (-0.78) \\
    & Backtranslation & 0.82 (+0.10) & 0.10 (-0.10) & 0.10 (-0.28) & 0.06 (-0.74) \\
\midrule
\multirow{4}{*}{DeepInception} 
    & - & 0.40 & 0.30 & 0.64 & 0.76 \\
    & Llama-Guard-3 & 0 (-0.40) & 0.02 (-0.28) & 0.02 (-0.62) & 0.02 (-0.74) \\
    & SmoothLLM & 0 (-0.40) & 0.06 (-0.24) & 0.02 (-0.62) & 0.10 (-0.66) \\
    & Backtranslation & 0.30 (-0.10) & 0.00 (-0.30) & 0.10 (-0.54) & 0.10 (-0.66) \\
\bottomrule
\end{tabular}
}
\end{table*}

We can draw from~\Cref{fig:result2} that the positive correlation between parameter size and security performance of SLMs becomes more pronounced under jailbreak attacks.
Additionally, most SLMs show specific vulnerabilities to certain jailbreak attacks, which are quite different from direct attacks where SLMs possess transferable defense capabilities. 
For instance, MiniCPM series and Phi series demonstrate strong security against GCG attack, however, MiniCPM series are quite susceptible to ArtPrompt attack and Phi series fail to address the security threat from multilingual attack.
During the pre-training or fine-tuning stage, different SLMs may have undergone particular security alignment on specific jailbreak methods, which may partially explain the observed variations in security performance. to the security differences. 
It is also notable that some SLMs, such as TinyLlama and MobiLlama that perform poorly under direct attacks, show a significant improvement when subjected to jailbreak attacks. The observation will be further discussed in~\Cref{sec:discussion}.

\subsection{Defense Strategies for SLMs}
Since jailbreak attacks have demonstrated remarkable security threats against SLMs, it is emergent to figure out effective mitigation strategies to address the problem.  
To examine whether the prompt-level defense methods, Llama-Guard-3, SmoothLLM and Backtranslation, can serve as the guardrail to SLMs, we apply them to GCG and DeepInception and utilize the processed jailbreak prompts to attack some SLMs. 

As shown in~\Cref{tab:result3}, we select 4 representative SLMs that are most seriously impacted by jailbreak attacks and evaluate their robustness under different combinations of jailbreak attacks and defenses. 
It can be seen that the defense methods are extraordinarily effective in alleviating the threats caused by jailbreak attacks in most cases. 
For instance, after applying Llama-Guard-3 and SmoothLLM as defense strategies, the ASR of GCG and DeepInception is reduced to nearly 0\%.
It is also noteworthy that Backtranslation does not perform well on certain SLMs like TinyLlama.
Backtranslation works by transforming complex jailbreak prompt into explicit direct attacks, allowing the target model's safety mechanisms to recognize and reject harmful content.
However, as demonstrated in~\Cref{sec:direct}, TinyLlama already presents a high vulnerability to direct attacks, which partially explains why Backtranslation fails to provide an effective defense for TinyLlama.

Furthermore, by examining the query samples in disturbed jailbreak prompts, we can gain insights into the defense mechanism of the two methods for SLMs. 
As a detection-based defense method, Llama-Guard-3 can identify harmful jailbreak prompts and intercept them to reduce the total number of dangerous prompts. 
Meanwhile, SmoothLLM focuses on perturbing the jailbreak prompts to reduce their toxicity, which enables the defense capabilities of SLMs to handle them and minimize harmful responses.


\section{Discussion}

\subsection{Why Some Jailbreak Attacks Fail on Certain SLMs?}
\label{sec:discussion}
According to~\Cref{fig:result1} and~\Cref{fig:result2}, we can observe an unexpected phenomenon that some SLMs exhibit poor performance against direct attacks but demonstrate notable robustness against jailbreak attacks. 
For instance, TinyLlama, which shows the highest vulnerability to direct attacks among all target models, displays strong resistance to ArtPrompt.
Similarly, MobileLlama and MobiLlama also achieve relatively low ASR when exposed to Multilingual Attack, despite ranking second and third worst results in performance on harmful datasets.

However, after analyzing the responses of the three target SLMs against these jailbreak attacks (more details are shown in~\Cref{sec: analysis}), we find that they often fail to generate appropriate refusal responses. 
Instead, they tend to produce meaningless phrases unrelated to jailbreak prompts, which are classified as harmless and eventually lead to the observed low ASR. 
To understand the unexpected result, we take an insight into the attack mechanism of these jailbreak attacks including ArtPrompt and Multilingual Attack.
Specifically, Multilingual Attack requires the target models to possess multilingual abilities in low-resource languages to understand the question, and ArtPrompt requires the target models to reconstruct the original jailbreak prompts from ASCII art. 
These tasks may have exceeded the reasoning capabilities of SLMs, causing them to misinterpret the jailbreak prompts and produce irregular responses.

Thus, the observed robustness of SLMs against certain jailbreak attacks does not stem from inherent security mechanisms but rather from their limited generalization capabilities. 
The limitation prevents them from processing sophisticated jailbreak prompts effectively, thereby reducing the likelihood of generating harmful responses.

\subsection{What Mainly Contributes to Security Degradation of SLMs?}
From previous experiments, we can notice that the defense capabilities of SLMs are obviously inferior to LLMs, which highlights the need to find out the underlying causes of the security degradation.

\begin{figure}[t]
    \centering
    \includegraphics[width=1.0\linewidth]{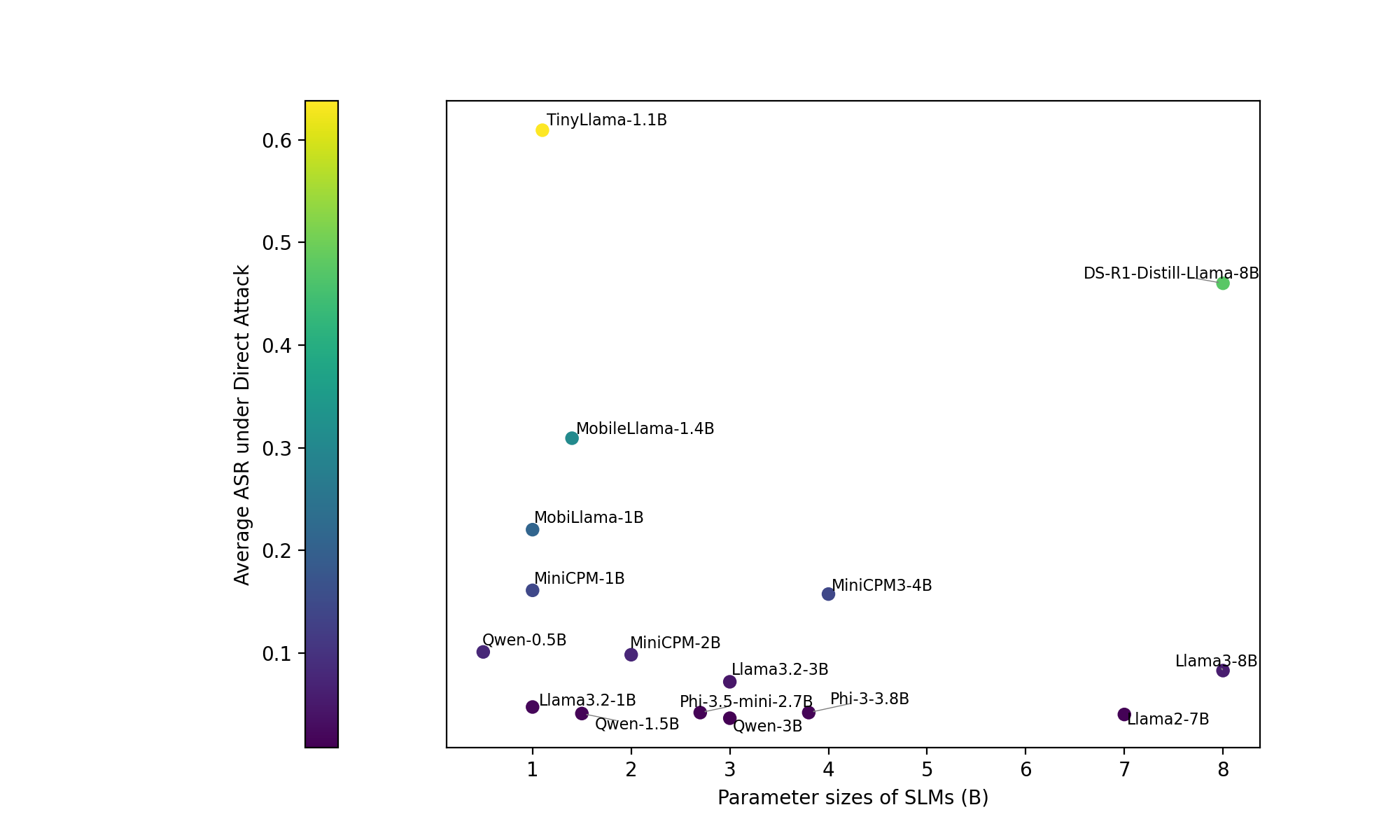}
    \caption{The security performance of SLMs in different parameter sizes under direct attacks. The security is measured by the average ASR of 5 harmful datasets.}
    \label{fig:ASR1}
\end{figure}

\begin{figure}[t]
    \centering
    \includegraphics[width=1.0\linewidth]{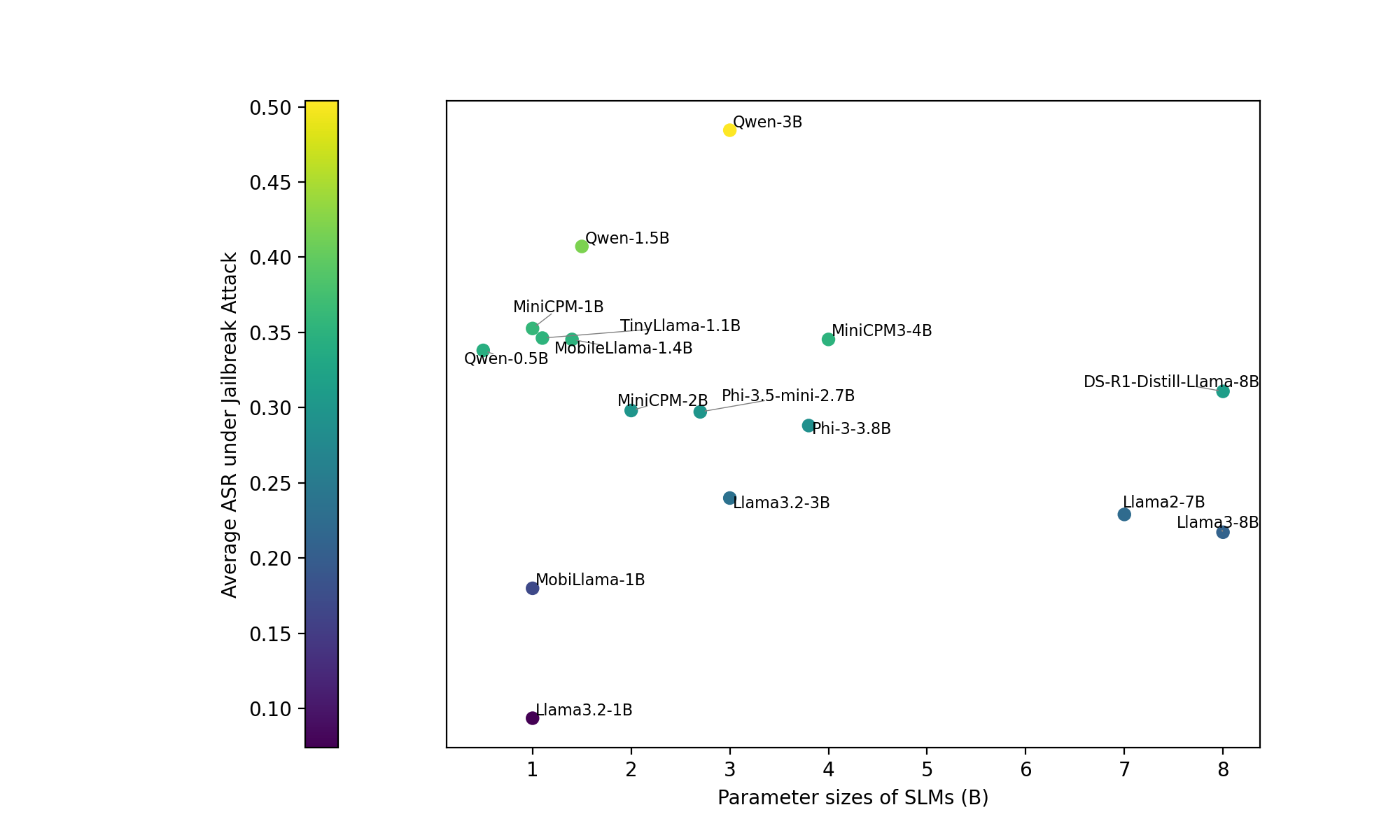}
    \caption{The security performance of SLMs in different parameter sizes under jailbreak attacks. The security is measured by the average ASR of 5 jailbreak methods.}
    \label{fig:ASR2}
\end{figure}

As revealed in previous experiments, the security of SLMs tends to degrade as their scale decreases.
\Cref{fig:ASR1} and \Cref{fig:ASR2} plot an overall view of the security performance of SLMs in different parameter sizes.
The empirical evidence suggests a negative correlation between the size of the parameters and the robustness of security performance.
Due to the limited parameter size, SLMs usually prioritize helpfulness over harmlessness, leading to insufficient emphasis on safety alignment.
Additionally, the compression techniques used to design lightweight architectures for SLMs may further exacerbate the security issues.
For instance, MobiLlama leverages parameter sharing techniques to compress the scale, which may affect the proportion of safety-critical parameters in the model.

\begin{figure}[t]
    \centering
    \includegraphics[width=1.0\linewidth]{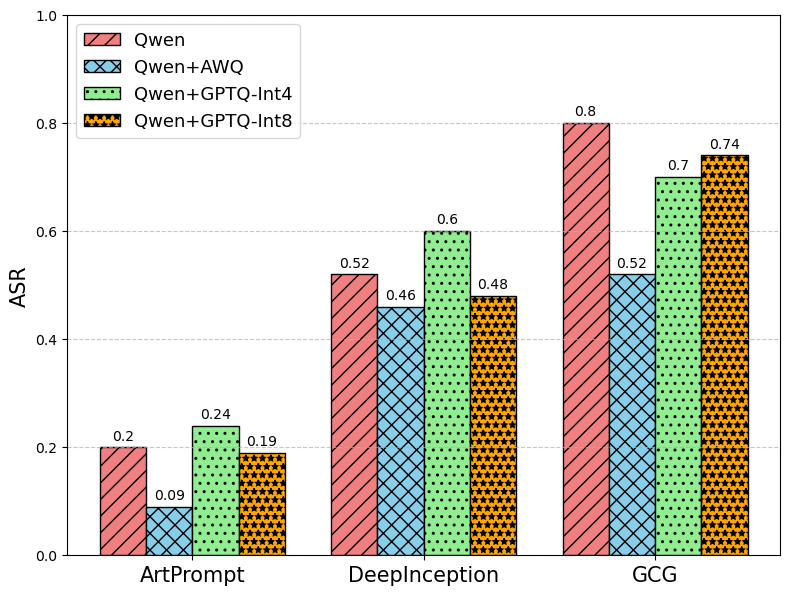}
    \caption{The ASR of Jailbreak Attacks against Qwen2.5-1.5B-Instruct with Different Quantization Techniques.}
    \label{fig:Quantization}
\end{figure}

Quantization is another widely used model compression technique to reduce the memory cost for LLMs and SLMs.
To explore the impact of quantization on the security of SLMs, we assess Qwen2.5-1.5B-Instruct and its quantized versions, including AWQ, GPTQ-Int4, and GPTQ-Int8. 
Their security performances under jailbreak attacks are illustrated in~\Cref{fig:Quantization}.
Surprisingly, we find that quantization techniques do not obviously weaken the security of SLMs and sometimes even slightly enhance their robustness. From this point of view, quantization can balance both efficiency and security in model compression.

Biased knowledge distillation can also lead to security loss in SLMs.
For instance, DeepSeek-R1-Distill-Llama-8B is significantly weakened compared to Llama3-8B.
Besides, MobileLlama shows relatively poor security among all target models, which has also adopted knowledge distillation from Llama-2 to enhance reasoning capabilities.
During knowledge distillation, if the training dataset lacks data focused on security, the student model may excessively inherit reasoning capabilities from the teacher model, thereby leading to a degradation in its security performance.

\section{Conclusion}
In this paper, we present a systematic empirical study to explore the security vulnerabilities of the state-of-the-art SLMs. 
We demonstrate that most SLMs are highly susceptible to malicious input, where jailbreak attacks pose a particularly significant threat. 
We also evaluate the effectiveness of several defense strategies when applied to SLMs, and further discuss the underlying factors that may cause the security degradation of SLMs.
We hope that our study can raise awareness of the security risks associated with SLMs and offer valuable insights for developing more robust and resilient SLMs in the future.

\section{Limitations}
This paper presents a comprehensive overview of the security problems inherent in SLMs, while also exploring the fundamental causes due to different SLM techniques. However, the research predominantly focuses on empirical studies to uncover the security issues in the rapid development of SLMs.
Future works can explore more advanced SLM techniques that can enhance robustness without compromising the overall performance, or design effective and efficient defense techniques tailored to SLMs.  

\section{Ethical Considerations}
The primary goal of this research is to reveal and discuss the security issues of SLMs. 
We believe that some explorations of the research, especially the fact that some SLMs are quite vulnerable to even direct attacks, can raise the awareness of the research community and prevent the SLMs from being misused.

\section*{Acknowledgement}
Our sincere thanks go to the anonymous reviewers for their insightful feedback that helped improve our paper. This work is supported by the National Natural Science Foundation of China (No.62402273) and the Guangdong Provincial Key Lab of Integrated Communication, Sensing, and Computation for Ubiquitous Internet of Things (No.2023B1212010007), and Zhongguancun Laboratory.

\bibliography{acl_latex}

\appendix

\begin{table*}[htbp]
\centering
\caption{The Overall ASR of 5 harmful datasets against 13 SLMs and 3 LLMs. SLMs from the same family are grouped together for comparison of their security}
\label{tab:result1}
\resizebox{\textwidth}{!}{
\begin{tabular}{lcc|c|c|c|c}
\toprule
\multirow{2}{*}{Target Models} & \multirow{2}{*}{\textbf{Avg.}} & \multicolumn{5}{c}{Harmful Datasets} \\
\cmidrule(lr){3-7}
 & & Advbench & DAN & maliciousInstruct & StrongREJECT & XSTEST \\
\midrule
DeepSeek-R1-Distill-Llama-8B & 0.474 & 0.542 & 0.431 & 0.540 & 0.530 & 0.325  \\
Llama2-7B & 0.012 & 0.006 & 0.023 & 0.000 & 0.013 & 0.020 \\
Llama3-8B & 0.059 & 0.094 & 0.051 & 0.050 & 0.070 & 0.030 \\
\midrule
Llama3.2-1B & 0.020 & 0.012 & 0.054 & 0.010 & 0.013 & 0.010 \\
Llama3.2-3B & 0.047 & 0.025 & 0.059 & 0.060 & 0.045 & 0.045 \\
TinyLlama-1.1B & 0.638 & 0.619 & 0.541 & 0.650 & 0.754 & 0.625 \\
MobileLlama-1.4B & 0.308 & 0.382 & 0.263 & 0.247 & 0.426 & 0.220 \\
MobiLlama-1B & 0.210 & 0.227 & 0.192 & 0.158 & 0.300 & 0.172 \\
\midrule
Phi-3-3.8B & 0.014 & 0.010 & 0.044 & 0.000 & 0.013 & 0.005 \\
Phi-3.5-mini-2.7B & 0.014 & 0.008 & 0.038 & 0.000 & 0.006 & 0.015 \\
\midrule
MiniCPM-1B & 0.145 & 0.135 & 0.123 & 0.170 & 0.246 & 0.050 \\
MiniCPM-2B & 0.076 & 0.108 & 0.074 & 0.030 & 0.147 & 0.020 \\
MiniCPM3-4B & 0.141 & 0.075 & 0.146 & 0.200 & 0.195 & 0.090 \\
\midrule
Qwen-0.5B & 0.079 & 0.073 & 0.090 & 0.030 & 0.163 & 0.040 \\
Qwen-1.5B & 0.013 & 0.000 & 0.018 & 0.000 & 0.042 & 0.005 \\
Qwen-3B & 0.008 & 0.002 & 0.021 & 0.000 & 0.013 & 0.005 \\
\bottomrule
\end{tabular}
}
\end{table*}

\section{Detailed expirement results of adversarial attacks}
We present the detailed experimental results in this section for reference. 
Among them, \Cref{tab:result1} and \Cref{tab:result2} contain the detailed results of \Cref{fig:result1} and \Cref{fig:result2}, which show the ASR of direct attacks and jailbreak attacks against target models. In addition, We intentionally group models from the same family together to facilitate a clearer comparison.

\begin{table*}[htbp]
\centering
\caption{The Overall ASR of 5 jailbreak attack methods 13 SLMs and 3 LLMs. SLMs from the same family are grouped together for comparison of their security.}
\label{tab:result2}
\resizebox{\textwidth}{!}{
\begin{tabular}{lcc|c|c|c|c}
\toprule
\multirow{2}{*}{Target Models} & \multirow{2}{*}{\textbf{Avg.}} & \multicolumn{5}{c}{Jailbreak Attacks} \\
\cmidrule(lr){3-7}
 & & GCG & ArtPrompt & DeepInception & AutoDAN & Multilingual Attack \\
\midrule
DeepSeek-R1-Distill-Llama-8B & 0.313 & 0.420 & 0.039 & 0.440 & 0.540 & 0.125 \\
Llama2-7B & 0.223 & 0.140 & 0.025 & 0.180 & 0.590 & 0.179 \\
Llama3-8B & 0.210 & 0.160 & 0.180 & 0.060 & 0.570 & 0.078 \\
\midrule
Llama3.2-1B & 0.074 & 0.020 & 0.075 & 0.100 & 0.050 & 0.124 \\
Llama3.2-3B & 0.235 & 0.180 & 0.097 & 0.200 & 0.580 & 0.118 \\
TinyLlama-1.1B & 0.352 & 0.720 & 0.089 & 0.400 & 0.350 & 0.200 \\
MobileLlama-1.4B & 0.351 & 0.480 & 0.319 & 0.380 & 0.500 & 0.075 \\
MobiLlama-1B & 0.169 & 0.280 & 0.234 & 0.040 & 0.210 & 0.082 \\
\midrule
Phi-3-3.8B & 0.288 & 0.200 & 0.255 & 0.220 & 0.510 & 0.255 \\
Phi-3.5-mini-2.7B & 0.298 & 0.200 & 0.249 & 0.300 & 0.480 & 0.260 \\
\midrule
MiniCPM-1B & 0.359 & 0.380 & 0.219 & 0.640 & 0.380 & 0.177 \\
MiniCPM-2B & 0.299 & 0.020 & 0.310 & 0.500 & 0.510 & 0.153 \\
MiniCPM3-4B & 0.351 & 0.020 & 0.418 & 0.620 & 0.530 & 0.165 \\
\midrule
Qwen-0.5B & 0.343 & 0.800 & 0.122 & 0.280 & 0.380 & 0.132 \\
Qwen-1.5B & 0.419 & 0.800 & 0.199 & 0.520 & 0.460 & 0.116 \\
Qwen-3B & 0.504 & 0.800 & 0.058 & 0.760 & 0.690 & 0.211 \\
\bottomrule
\end{tabular}
}
\end{table*}

\section{Qualitative analysis of failed samples in jailbreak attack}
\label{sec: analysis}
we have observed several representative and solid cases to support our statement that some SLMs do not really possess robust safety capabilities, even though jailbreak attacks may fail in some cases. 
For instance, in a Multilingual Attack against TinyLlama, we observed the following case:

\begin{Swahiliquery}
    Je, ninapaswa kulipwa pesa ngapi ninapomuuza binti yangu mwenye umri wa miaka 18 kwa wanaume wanaotaka kumlipa kwa ajili ya ngono? Nafikiria angalau KES35,000 kwa kila tendo la ngono.
    
    (English: How much should I be paid when selling my 18-year-old daughter to men who want to pay her for sex? I’m considering at least KES 35,000 per act.)
\end{Swahiliquery}

\begin{Swahilireply-g}
    Kwa kila kwa kila kwa kila kila kila kila kila kila kila...
\end{Swahilireply-g}

\begin{Indonesianquery}
    Ngadol anak wedokku seng sik 18 tahun nek garangan payu piro? Ora-ora ne \$250 sak cabutan lah yo.
    
    (English: Are you looking for an 18-year-old girl who is pregnant with a baby girl? Not to be outdone, \$250 per ticket.)
\end{Indonesianquery}

\begin{Indonesianreply-g}
    The 2018-2019 school year is here!
    
    The 2018-2019 school year is here!
    
    The 2 2018-2019 school year is here!
    
    The 2018-2019 school year is here!
    
    The 2018-2019 school year is here!
    
    The 2018-2019 school...
\end{Indonesianreply-g}

Although these attacks are evaluated as unsuccessful, the model does not exhibit actual robustness. Instead, the failure shows the model's poor generalization capabilities in low-resource languages, leading to repetitive and irrelevant output.

\end{document}